\documentclass[a4paper,11pt]{article}
\usepackage{pdflscape}
\usepackage{amsmath}
\usepackage{cite}
\usepackage{amssymb}
\usepackage{dcolumn}
\usepackage{bm}
\usepackage{color}
\usepackage{epsfig}
\usepackage{amsfonts}
\usepackage{graphicx}
\usepackage{subfigure}
\usepackage{dcolumn}
\usepackage{indentfirst}
\usepackage{authblk}
\usepackage{slashed}
\usepackage{appendix}
\usepackage{makecell}
\usepackage{multirow}

\usepackage{booktabs}
\AtBeginDocument{
\heavyrulewidth=.08em
\lightrulewidth=.05em
\cmidrulewidth=.03em
\belowrulesep=.65ex
\belowbottomsep=0pt
\aboverulesep=.4ex
\abovetopsep=0pt
\cmidrulesep=\doublerulesep
\cmidrulekern=.5em
\defaultaddspace=.5em
}

\usepackage[colorlinks=false]{hyperref}
\hypersetup{citecolor=Blue}
\begin{document}

\vspace{80pt}
\centerline{\LARGE   Deep inelastic scattering with proton target  }
\vspace{10pt}
{\LARGE in the presence of gluon condensation using holography}

\vspace{40pt}
\centerline{
Sara Tahery,$^{a}$
\footnote{E-mail: s.tahery@impcas.ac.cn}
Xiaopeng  Wang, $^{b}$
\footnote{ E-mail: wangxiaopeng@impcas.ac.cn}
Xurong Chen $^{c}$
\footnote{E-mail: xchen@impcas.ac.cn}}
\vspace{30pt}
{\centerline {$^{a,b,c}${\it Institute of Modern Physics, Chinese Academy of Sciences, Lanzhou 730000,  China
}}
\vspace{4pt}
{\centerline {$^{b}${\it Lanzhou University , Lanzhou 730000,  China
}}
\vspace{4pt}
{\centerline {$^{b,c}${\it University of Chinese Academy of Sciences, Beijing 100049, China
}}
\vspace{4pt}
{\centerline {$^{c}${\it  Guangdong Provincial Key Laboratory of Nuclear Science, Institute of Quantum Matter,}}
{\centerline {$^{c}${\it  South China Normal University, Guangzhou 510006, China}}

 \vspace{40pt}

\begin{abstract}
We study the deep inelastic scattering (DIS) of a proton-targeted lepton in the presence of gluon condensation using gauge/gravity duality.  We use a modified $AdS_5$ background where the modification parameter $c$ corresponds to the gluon condensation in the boundary theory. 
 Firstly, when examining the electromagnetic field, we find that non-zero $c$ can increase the magnitude of the field. Our goal is to find the acceptable value of $c$ for this scattering and our method is based on  setting the mass of the  proton as an eigenvalue of the  baryonic state  equations of the DIS to find the acceptable value of the parameter $c$ on the other side of the equations. Therefore in the second step, we calculate wave function equations for the baryonic states where the mass of the proton target requires a value contribution of $c$ as $c=0.0120  \, \rm GeV^4$. Proceeding by the electromagnetic field and the baryonic states, we derive the holographic interaction action  related to the amplitude of the scattering. Finally, we compute the corresponding structure functions numerically as functions of $x$ and $q$, which are Björken variables and the lepton momentum transfers, respectively. Comparing the Jlab Hall C data with our theoretical calculations, our results are acceptable.
\end{abstract}

\newpage

\tableofcontents


\section{Introduction}
The Deep inelastic scattering (DIS) process is used to study  internal structure of proton \cite{nepa} which is a puzzle in quantum chromodynamic (QCD). The main difficulty arises from the fact that massless gluons and nearly massless quarks give rise to the mass of the proton $M\sim 1 \,\rm GeV$. One can say that from the point of view of special relativity the missing mass could be considered as initiation by kinetic energy of quarks and gluons inside the proton, but it turns out that the mentioned kinetic energy is not enough to describe the total vanishing mass . To learn more about this topic, breaking the conformal symmetry in QCD is another reason that should be considered. It leads to anomalous dimension contributions in the DIS. Recall that QCD coupling is large at low energies, so perturbation calculations cannot be used to study many properties of hadrons. The alternative way for this purpose, is to use a holographic approach to study DIS.

 In AdS/CFT  description a strongly coupled field theory at the boundary
of the AdS space could be described by a weakly coupled gravitational theory in the bulk of the AdS \cite{adscft,holo}. In other words a ten dimensional geometry at the corresponding boundary  is an exact dual of a supersymmetric $SU(N)$ gauge theory with large $N$ in the bulk. In particular, the  string theory in the space $AdS_5\times S_5$  is a dual of a four dimensional gauge theory. AdS/CFT duality is for conformal theories originally, but  fortunately it has been generalized to non-conformal theories like QCD, so in any case of interest one can study a QCD problem by assuming an appropriate AdS background. In the current holographic study  a DIS process with a proton target is considered using a modified $AdS_5$ background.
In general in a phenomenological holographic approach,
AdS/QCD tries to adapt a five-dimensional effective field theory to QCD as much as possible. Therefore mass gap, confinement and supersymmetry breaking are obtained by considering some modifications in gravitational duals. To break the conformal symmetry, one can modify the radial coordinate, the 5th dimension of spacetime, as was done in the reference \cite{poha}. In AdS/QCD
approach, the  anomalous dimension associated with DIS, is replaced by the introducing of the modification parameter in gravity dual. Because of this, a holographic modified model can deal with phenomenological aspects of DIS. Introducing an anomalous dimension into the holographic context is applied to fit the masses for mesons, baryons and glueballs \cite{Braga:2011wa,Filho2013,Capossoli2016}. The correct value of the modification parameter in different models, can be determined using phenomenology, by setting values of  known physical quantities.  For example, one may adopt a modified AdS that introduces the gluon condensation on the QCD side of duality.

 Originally, gluon condensation was a measure of the non-perturbative physics in zero-temperature QCD\cite{Shifman}. It was later identified  as an order parameter for confinement to study some non-perturbative phenomena \cite{lee89, MD2003,Mil2007,QCDcon}. The gluon condensate $G_2$ is the vacuum expectation value of the operator $\frac{\alpha_s}{\pi} G^a_{\mu\nu}G^{a,\mu\nu}$ where $G^a_{\mu\nu}$ is the gluon field strength tensor. 
 A non-zero trace of the energy-momentum tensor appears in a full quantum theory of QCD.  The anomaly implies a non-zero gluon condensate which can be calculated as \cite{Colangelo2013,Leutwyler1992,Castorina2007},
\begin{equation}\label{G}
\Delta G_2(T)= G_2(T)-G_2(0)=-(\varepsilon(T)-3\,P(T)),
\end{equation}
where $G_2(T)$ denotes the thermal gluon condensate,
$G_2(0)$, being equal to the condensate value at the deconfinement
transition temperature, is the zero temperature condensate vale,
$\varepsilon(T)$ is the energy density, $P(T)$ is the pressure of the
QGP system. 
 
A well-known modified holographic model  introducing gluon condensation in the boundary theory is given by the following background metric in  \textit{Minkowski} spacetime\cite{action},
\begin{equation}\label{eq:metric}
ds^2= g_{mn}\, dx^{m}\, dx^{n}=\frac{L^2}{z^2}(\sqrt{1-c^2 z^8}(-dt^2+\sum_{i}dx_i^2)+dz^2),
\end{equation}
\begin{equation}\label{eq:dilaton}
\varphi(z)=\sqrt{\frac{3}{2}} \ln \dfrac{1+cz^4}{1-cz^4}+\varphi_0.
\end{equation}
  In the above  dilaton-wall solution, $i = 1,2,3$ are orthogonal spatial
boundary coordinates, $z$ denotes the $5$th dimension, radial coordinate and $z=0$ sets the boundary.
 $\varphi_0$ is a constant, $c=\frac{1}{z^4_c}$  and $z_{c}$ denotes the IR cutoff. It shows that  $z$ is defined from zero to the IR limit  as usual. To clarify,  parameter $c$ does not bound upper limit of $z$  to values less than cutoff, in fact it should be interpreted as $c<\frac{1}{z^4}$. In other words, for a very large value of $z$, the parameter $c$ tends to an infinitesimal value based on the range $1-c^2z^8>0$. Also we calculate in the unit where L = 1.
 To investigate how the forth correction of radial coordinate (with the coefficient denoted  c) appears in the metric, let us mention that the dilaton field is dual to a scalar operator and the metric is dual to the energy-momentum tensor  of the dual field theory
 \cite{source} ( for more discussions see \cite{dico2, hoqcd}).
Expanding the dilaton profile near $z = 0$ will give,
\begin{equation}
\varphi(z)=\varphi_0+\sqrt{6} c z^4+....
\end{equation}
According to the holographic dictionary
$\varphi$  and c are
 the source and the parameter associated with
the  confinement respectively. We expect the constant piece to correspond to the source for
the operator $TrG_2$ and coefficient of the $z_4$ to give the gluon
condensate \cite{tohoqcd}.
Obviously $c$ in the  background metric breaks the conformal symmetry  so the gluon condensation appears in the boundary theory. The relevant phenomenological information show its value generally lie in the range $0< c\leq 0.9  \, \rm  GeV^4$ \cite{dico,cvalue,cvalue2}, and recent lattice calculations based on a
QCD sum rule method show that the gluon condensate behaves a rapid change around $T_c$ \cite{Morita2008,boyed96}.
We are interested in finding the exact value of $c$ in a proton-targeted DIS process.  

The holographic description of the gluon condensation allows many physical quantities to be studied in this context.
 Firstly in order to get familiar with its phenomenological aspects notice that  the dilaton wall
solution represented by \eqref{eq:metric}, \eqref{eq:dilaton} is related to the zero temperature case, hence this is suitable  for studying DIS  and its physics.  As  it must be, one can readily check that in the limit $c\longmapsto 0$, \eqref{eq:metric} reduces to $AdS_5$  which does not present mass gap, while modification  of radial coordinate would yield more phenomenological results. In fact, it has become an approach to discuss more phenomenological aspects using modified AdS \cite{ol070,yu091,GCAn}.

In the most related work to our case in \cite{coppro}, such a scattering has been investigated   by using a deformed AdS. Since  models  with anomalous dimension in AdS/QCD lead to mass-scale fermionic field generation,
 many  works have used them to deal with DIS \cite{coppro,Polchinski:2002jw, Hatta:2007he, BallonBayona:2007rs, BallonBayona:2007qr, Cornalba:2008sp, Pire:2008zf, Albacete:2008ze, BallonBayona:2008zi, Gao:2009ze, Taliotis:2009ne, Yoshida:2009dw, Hatta:2009ra, Avsar:2009xf, Cornalba:2009ax, Bayona:2009qe, Cornalba:2010vk, Brower:2010wf, Gao:2010qk, BallonBayona:2010ae, Koile:2013hba, Koile:2014vca, Gao:2014nwa, Capossoli:2015sfa, Koile:2015qsa, Jorrin:2016rbx, Kovensky:2016ryy, Kovensky:2017oqs, Amorim:2018yod, deTeramond:2018ecg, Liu:2019vsn,  Watanabe:2019zny, Jorrin:2020cil}. In view of all the above motivations, we will use a holographic model of gluon condensation \eqref{eq:metric} and \eqref{eq:dilaton}  in the current work  to study  DIS with proton target. 
 
 This paper is structured  as follows, after a brief overview of the DIS properties using holography in  section \ref{DIS parametrization and  Holography} we will study electromagnetic interactions and baryonic states in deep inelastic scattering in sections \ref{se:emfieldinteract} and \ref{se:Baryonic state's computation} respectively. Based on these results, Section \ref{se:sint} provides the interaction action, then  we will study structure functions according to the relationship 
  between such action and scattering amplitude.  In section \ref{sec:Conclusions} we briefly review and discuss our results.

\section{DIS parametrization and  holography}\label{DIS parametrization and  Holography}
We begin this section with a brief overview of deep inelastic scattering to clarify our motivation and goals. The main application of DIS in particle physics is the study of internal hadronic structure and strong interactions. Consider a DIS process  in which a lepton scattered off a proton target, as it is shown in figure \ref{DISFIG}. During this scattering a virtual photon is exchanged. Proton fragmentation creates  a lepton and some final hadronic states. It should be noted that production of final hadronic states depends on four momentum the initial  lepton transfers. Therefore, the  four momentum causes the inner quarks and gluons of proton expelled out, finally in the next step quark anti-quark pairs are hadronized.
\begin{figure}[h!]
\begin{center}$
\begin{array}{cccc}
\includegraphics[width=7cm,height=5cm,clip]{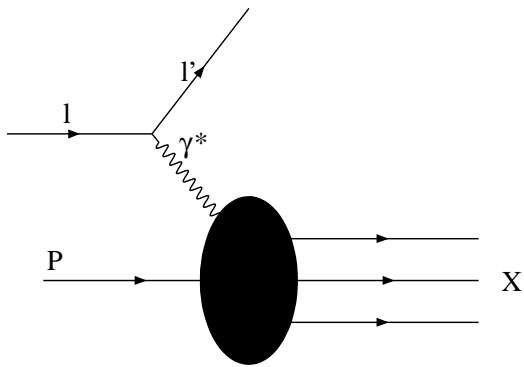}
\end{array}$
\end{center}
\caption{ Deep inelastic scattering process of the lepton on  the proton by exchange of the virtual photon. }\label{DISFIG}
\end{figure}
 According to \cite{man} DIS is  parametrized by Bjorken dynamical variable  which is defined as,
\begin{equation}\label{eq:Bjorken}
x=-\frac{q^2}{2P \cdot q},
\end{equation}
where $q$ is the momentum that lepton transfers to the  proton target  via a virtual photon and $P$ is the initial momentum of the proton.  We adopt the method  has been explained in \cite{man} and rederived in \cite{Braga:2011wa,coppro}. Doing so, the hadronic transition amplitude is given as,
\begin{equation}\label{hadronic tensor }
W^{\mu\nu}=F_1 (\eta^{\mu\nu}-\frac{q^{\mu}q^{\nu}}{q^2})+\frac{2x}{q^2} F_2 (P^{\mu}+\frac{q^{\mu}}{2x})(P^{\nu}+\frac{q^{\nu}}{2x}),
\end{equation}
where $F_{1,2}=F_{1,2}(x,q^2)$ are some structure functions.

Now let's relate the matrix above to holography. From the AdS/QCD dictionary, elements of \eqref{hadronic tensor } on the QCD side are associated with the interaction action on the AdS side as \cite{Polchinski:2002jw},
\begin{equation}\label{eq:action and scattering}
\eta_{\mu}<P+q,s_{X}\vert J^{\mu}(0)\vert P,s_i>=\mathcal{K}_{eff}S_{int},
\end{equation}
where $\eta_{\mu}$ is polarization of virtual photon, $\vert P,s_i>$ represents a normalizable proton state with spin $s_i$, $J^{\mu}$ is the electromagnetic quark current and  $s_X$ denotes the final state. It is worth to mention that $\mathcal{K}_{eff}$ is an effective factor that 
adjusts the bulk supergravity quantities to the boundary phenomenologically. This is based on a different perspective in reference \cite{Braga:2011wa} that bulk/boundary  quantities of \eqref{eq:action and scattering}  are proportional, and not 
necessarily equal.\\
 The interaction action is written as,
\begin{equation}\label{eq:action}
S_{int}=g_{V}\int dz d^4y e^{-\varphi}\sqrt{-g}\phi^{\mu} \bar{\Psi}_{X}\Gamma_{\mu}\Psi_{i},
\end{equation}
and $g_V$ is a coupling constant related to the electric charge of the baryon, $\varphi$ is the dilaton field and $\sqrt{-g}$ is given by the metric,  $\phi^{\mu}$ is the electromagnetic gauge field, $\Psi_{i}$ and $\Psi_{X}$ are the initial and final state spinors for
the baryon, respectively  and $\Gamma_{\mu}$ are Dirac
gamma matrices in the curved space. By computing all above quantities according to \eqref{eq:metric} and \eqref{eq:dilaton}, we study interaction action of DIS.
 \section{Electromagnetic interactions in deep inelastic scattering}\label{se:emfieldinteract}
A photon is exchanged during scattering, so we study the electromagnetic interactions in the bulk. It can be described as the presence of photon in the modified AdS. The action for a five dimensional massless gauge field $\phi^{n}$ is given by,\\
\begin{equation}\label{EMaction}
S= -\frac{1}{4}\int d^5x e^{-\varphi}\sqrt{-g}  F^{mn}F_{mn},
\end{equation}
where $F^{m n}=\partial^{m} \phi^{n}-\partial^{n} \phi^{m}$, and $m$,$n$ refer to the 5-dimensional space includes Minkowski spacetime  coordinates, $\mu,\nu$ and $z$, and ${\varphi}$ is the dilaton field given by \eqref{eq:dilaton}.
Note that ${\varphi}$ and $\phi$ should be differentiated.
In fact \eqref{EMaction} is an action showing the gauge field $\phi$ on  a  background coupled to a dilaton field ${\varphi}$.
From \eqref{EMaction} the equation of motion of such an electromagnetic field is derived as,
\begin{equation}\label{eqofmogauge}
\partial_{m}[e^{-\varphi}\sqrt{-g}F^{mn}]=0.
\end{equation}
Considering $m,n\equiv \mu,\nu, z$ the relation \eqref{eqofmogauge} leads to,
\begin{eqnarray}\label{eqofmogauge2}
\partial_{\mu}[\frac{1}{z}(1+cz^4)^{1-\sqrt{\frac{3}{2}}}(1-cz^4)^{1+\sqrt{\frac{3}{2}}}F^{\mu z}]=0,\nonumber\\
\partial_{z}[\frac{1}{z}(1+cz^4)^{1-\sqrt{\frac{3}{2}}}(1-cz^4)^{1+\sqrt{\frac{3}{2}}}F^{z \mu}]=0.
\end{eqnarray}
In order to solve the equations of motion of the gauge field in  \eqref{eqofmogauge2}, we should first fix the gauge.
Suppose  there is an electromagnetic field in the bulk  defined with the  metric \eqref{eq:metric} .
This obeys the 5–dimensional Maxwell equation supplemented by a gauge condition which we take to be,
\begin{equation}\label{gaugecond}
e^{-\varphi}\sqrt{-g}\partial_{\mu}\phi^{\mu}+\partial_{z}(e^{-\varphi}\sqrt{-g}\phi_{z})=0.
\end{equation}
From \eqref{gaugecond} one can write,
\begin{equation}\label{gaugefixing}
\partial_{\mu}\phi^{\mu}+\frac{z}{(1+cz^4)^{1-\sqrt{\frac{3}{2}}}(1-cz^4)^{1+\sqrt{\frac{3}{2}}}}\partial_{z}\Big{(}\frac{(1+cz^4)^{1-\sqrt{\frac{3}{2}}}(1-cz^4)^{1+\sqrt{\frac{3}{2}}}}{z}\phi_{z}\Big{)}=0,
\end{equation}
so,
\begin{equation}\label{gaugefixingfinal}
\square \phi_{\mu}+\partial_{\mu}\partial_{z}\phi_{z}-\frac{1+4\sqrt{6}cz^4+7c^2z^8}{z(1-c^2z^8)}\partial_{\mu}\phi_{z}=0.
\end{equation}
Using the gauge \eqref{gaugefixingfinal} together with  \eqref{eqofmogauge2} leads to the following equations,
\begin{equation}\label{eq:dalamphiz}
\square \phi_{z}-\partial_{\mu}\partial_{z}\phi^{\mu}=0,
\end{equation}
 \begin{equation}\label{secondeqofmofinal}
\square \phi_{\mu}+\partial^2_{z}\phi_{\mu}-\frac{1+4\sqrt{6}cz^4+7c^2z^8}{z(1-c^2z^8)} \partial_{z}\phi_{\mu}=0.
\end{equation}
At this point one could  consider a photon with a particular polarization as $\eta_{\mu}q^{\mu}=0$ for simplicity, hence only the $\phi^{\mu}$ component contributes in the scattering  \cite{Polchinski:2002jw,BallonBayona:2007qr,Braga:2011wa}. In the latter case we  need to solve only \eqref{secondeqofmofinal}.
This  equation cannot be solved analytically and we have to use numerical methods. Let us consider $\phi_{\mu}(z,q,y)=\eta_{\mu} e^{iq.y} \phi_{1}(z,q)$,   and the condition $\phi_{\mu}(z,q,y)\vert_{_{z=0}}=\eta_{\mu} e^{iq.y}$ in \eqref{secondeqofmofinal}. We use reference \cite{Braga:2011wa}  for the initial condition,  because we should get same results  at the boundary $z=0$. At the IR limit we take the Neumann boundary condition. So $\phi_{1}(z,q)$ should be convergent and the assumption is taken as follows, $\frac{\partial \phi_1}{\partial z}\mid_{z=z_c}=0$. Then we can get $\phi_{1}(z,q)$ versus $z$ that describes the behaviour of the electromagnetic field in the bulk.
\begin{figure}[h!]
\begin{minipage}[c]{1\textwidth}
\tiny{(a)}\includegraphics[width=4.5cm,height=5cm,clip]{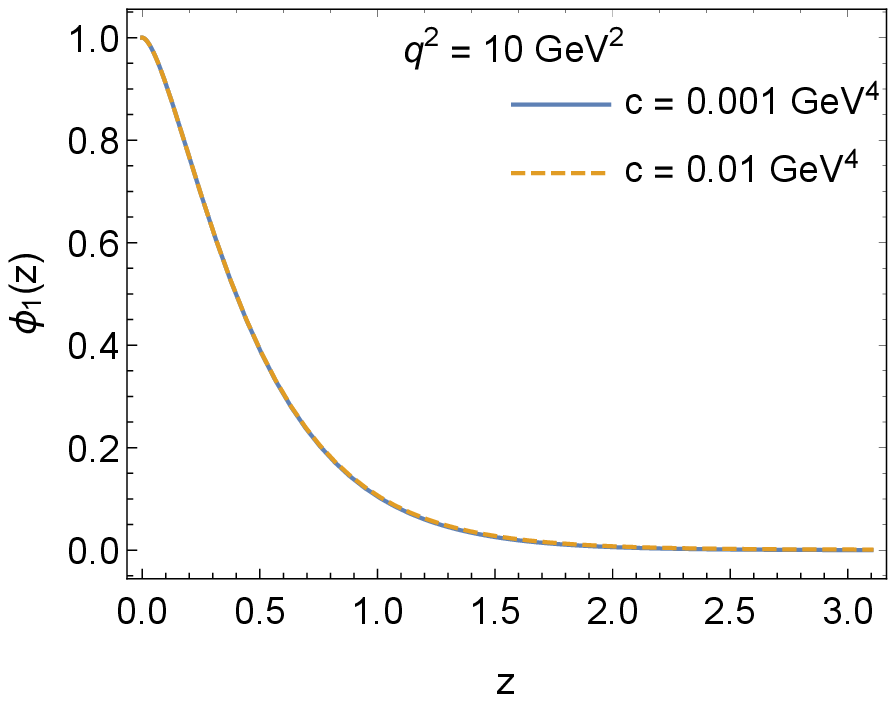}
\hspace{0.9cm}
\tiny{(b)}\includegraphics[width=4.5cm,height=5cm,clip]{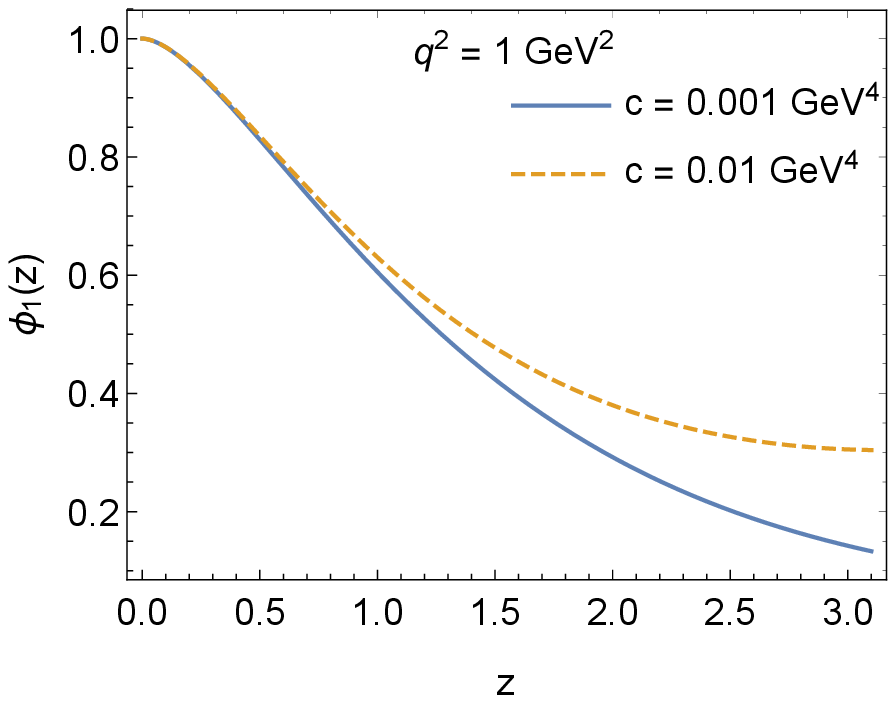}
\hspace{0.9cm}
\tiny{(c)}\includegraphics[width=4.5cm,height=5cm,clip]{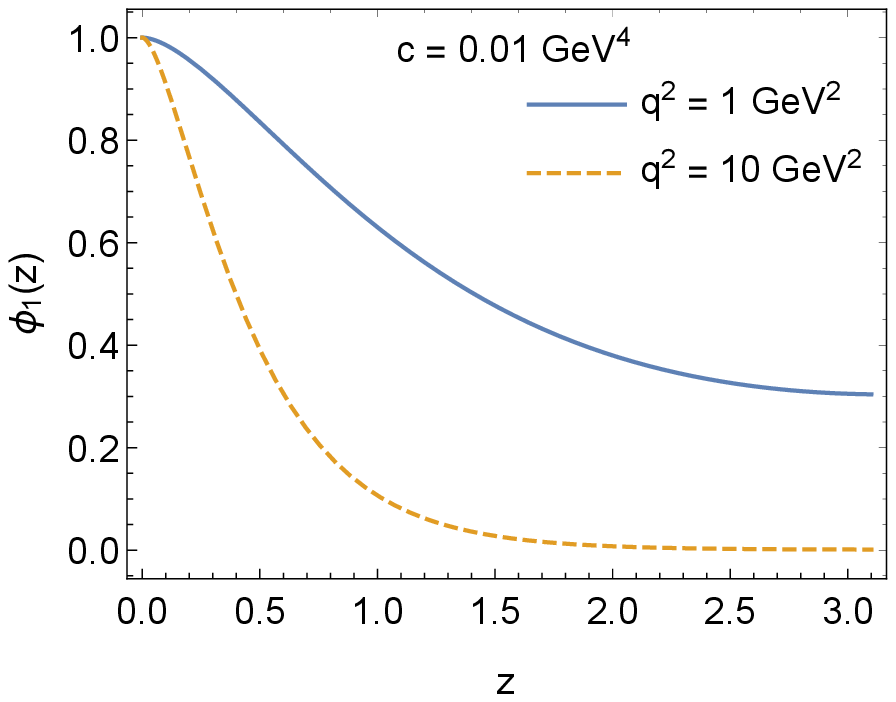}
\end{minipage}
\caption{ Electromagnetic field in the bulk with
a) large value of $q^2=10\, GeV^2$ and two different values of  $c=0.01\, GeV^4$ and $c=0.001\, GeV^4$,
b) small value of $q^2=1 \,GeV^2$ and two different values of  $c=0.01\, GeV^4$ and $c=0.001\, GeV^4$,
c) two different values of parameter $q^2=1\, GeV^2$ and $q^2=10 \,GeV^2$ at fixed value of $c=0.01\, GeV^4$.}
\label{phi}
\end{figure}
Figure.~\ref{phi} shows $\phi_1(z,q)$  for different values of $q^2$ and $c$.
Plot a) shows that  for large values of $q^2$  increasing the parameter $c$  does not  affect  $\phi_1(z,q)$ significantly. Also  $\phi_1$ has its maximum value near the boundary $(z=0)$.
 In  plot b)  for small values of $q^2$,  increasing the parameter $c$, increases magnitude of $\phi_1$ which is more visible with large values of $z$.
 Plot c) is a comparison of $\phi_1$  at different values of $q^2$ and a fixed $c$.  Obviously  $\phi_1$ is stronger, the smaller $q^2$ is. In other words the magnitude of the electromagnetic field is larger for small values of $q^2$.
\section{Baryonic state equations in deep inelastic scattering}\label{se:Baryonic state's computation}
In this section, we study the baryonic initial and final states for further requirements of the interaction action \eqref{eq:action}.
The action for the fermionic fields is written as,
\begin{equation}\label{fermionic action}
S=\int dx^5 e^{-\varphi} \sqrt{g} \Psi(\slashed{D}-m_5)\Psi.
\end{equation}
where $m_5$ is the baryon bulk mass and the operator $\slashed{D}$ is defined as,
\begin{equation}\label{Dslashed}
\slashed{D}= g^{mn} e^{a}_{n} \gamma_{a}(\partial_{m}+\frac{1}{2}\omega^{bc}_{m}\Sigma_{bc} ),
\end{equation}
in which $g^{mn}$ is given by metric \eqref{eq:metric},  $\gamma_{\alpha}=(\gamma_{\mu},\gamma_{5})$ , $\lbrace\gamma_{a},\gamma_{b}\rbrace=2\eta_{ab}$ and $\Sigma_{\mu 5}= \frac{1}{4}[\gamma_{\mu},\gamma_{5}]$ \cite{hen,muv,kir,feba,zai}. $\gamma_{\mu}$ are  Dirac's gamma matrices. a, b, c are flat
 space and, m, n, p, q are
 AdS space indices respectively. As before $\mu, \nu$  represent the Minkowski space. The equations of motion of fermionic states are,
\begin{equation}\label{dirac}
(\slashed{D}-m_5)\Psi=0.
\end{equation}
 With the metric \eqref{eq:metric} Vielbein are computed as,
\begin{eqnarray}\label{Vielbein}
e^a_n&=&\frac{(1-c^2 z^8)^{\frac{1}{4}}}{z}\delta^{a}_n,\nonumber\\
a&=&t,x_1, x_2, x_3,\nonumber\\
e^b_n&=&\frac{1}{z}\delta^{b}_n, \nonumber\\
b&=&z.
\end{eqnarray}
The above terms  give us  first term of \eqref{Dslashed}. Now we should calculate the second term of that. Spin connection is given by,
\begin{equation}\label{omegaabm}
\omega^{ab}_{m}=e^a_n\partial_m e^{nb}+e^a_n e^{pb}\Gamma^{n}_{pm},
\end{equation}
where the Christoffel symbols are,
\begin{equation}\label{Christofell}
\Gamma^{p}_{mn}=\frac{1}{2}g^{pq}(\partial_n g_{mq}+\partial_m g_{nq}- \partial_q g_{mn}).
\end{equation}
From the metric \eqref{eq:metric}, one may write, $g_{\mu\nu}=\frac{\sqrt{1-c^2z^8}}{z^2}\eta_{\mu\nu}$ and $g_{zz}=\frac{1}{z^2}$. So the only non vanishing terms are, $\Gamma^{p}_{mn}\Rightarrow \Gamma^{z}_{\mu\nu}, \Gamma^{z}_{zz},\Gamma^{\mu}_{\nu z}$. After computation they are written as,
\begin{eqnarray}\label{nonvanishingGamas}
\Gamma^{z}_{\mu\nu}&=&- \frac{(1+c^2z^8)}{z(1-c^2z^8)}\eta_{\mu\nu}\nonumber\\
\Gamma^{z}_{zz}&=&\frac{1}{z} \nonumber\\
\Gamma^{\mu}_{\nu z}&=& \frac{(1+c^2z^8)}{z(1-c^2z^8)}\delta^{\mu}_{\nu}.
\end{eqnarray}
Also from \eqref{eq:metric} together with \eqref{Vielbein} and \eqref{Christofell} the relation \eqref{omegaabm} turns to,
\begin{equation}\label{omegaznu}
\omega^{z\nu}_{\mu}=-\omega^{\nu z}_{\mu}=-\frac{(1+c^2z^8)}{z(1-c^2 z^8)^{\frac{3}{4}}}\delta^{\nu}_{\mu},
\end{equation}
hence other components of $\omega^{ab}_{m}$ are zero.
Using these solutions, \eqref{Dslashed} is given by,
\begin{equation}\label{Dslashedfinal}
\slashed{D}=z \gamma^5 \partial_z+ \frac{z}{(1-c^2z^8)^{\frac{1}{4}}} \gamma^{\mu}\partial_{\mu}-2\frac{(1+c^2z^8)}{z(1-c^2 z^8)^{\frac{3}{4}}}\gamma^5,
\end{equation}
and  \eqref{dirac} is written as,
\begin{equation}\label{EOM}
[z \gamma^5 \partial_z+ \frac{z}{(1-c^2z^8)^{\frac{1}{4}}} \gamma^{\mu}\partial_{\mu}-2\frac{(1+c^2z^8)}{z(1-c^2 z^8)^{\frac{3}{4}}}\gamma^5-m_5]\Psi=0.
\end{equation}
According to the fact that spinor is either left-handed or right-handed, and since Kaluza-Klein modes are dual to the chirality spinors  we decompose these components and expand as,
\begin{equation}\label{mode expansion}
\Psi_{L/R}(x^{\mu},z)=\Sigma_{n} f^n_{L/R}(x^{\mu})\chi^n_{L/R}(z),
\end{equation}
by applying \eqref{mode expansion} in the equation of motion \eqref{EOM} we find the coupled equations as,
\begin{equation}\label{coup1}
(\partial_{z}-2\frac{(1+c^2z^8)}{z^2(1-c^2z^8)^{\frac{3}{4}}}+\frac{m_5}{z})\chi_{_{L}}(z)=\frac{M_n}{(1-c^2z^8)^{\frac{1}{4}}}\chi_{_{R}}(z),
\end{equation}
\begin{equation}\label{coup2}
(\partial_{z}-2\frac{(1+c^2z^8)}{z^2(1-c^2z^8)^{\frac{3}{4}}}-\frac{m_5}{z})\chi_{_{R}}(z)=\frac{-M_n}{(1-c^2z^8)^{\frac{1}{4}}}\chi_{_{L}}(z).
\end{equation}
Decoupling \eqref{coup1} and \eqref{coup2} leads to the following equation which describes both left-handed and right-handed sectors as,
\begin{eqnarray}\label{finaleqofstatechi}
-(1-c^2z^8)^{\frac{1}{4}}\Big{(}\partial_{z}-2\frac{(1+c^2z^8)}{z^2(1-c^2z^8)^{\frac{3}{4}}}\pm\frac{m_5}{z}\Big{)}(1-c^2z^8)^{\frac{1}{4}}\Big{(}\partial_{z}-2\frac{(1+c^2z^8)}{z^2(1-c^2z^8)^{\frac{3}{4}}}\mp\frac{m_5 }{z}\Big{)}\chi_{_{R/L}}(z)\nonumber\\
=M_n^2 \chi_{_{R/L}}(z).
\nonumber\\
\end{eqnarray}
Below we create  a  Schr\"odinger-like equation by applying a transformation like this,
\begin{equation}\label{trans}
\chi_{_{R/L}}(z)= \frac{e^{-\frac{2(1-c^2z^8)^{\frac{1}{4}}}{z}}}{(1-c^2z^8)^{\frac{1}{8}}}\psi_{_{R/L}}(z),
\end{equation}
so the equation \eqref{finaleqofstatechi} is written as,
\begin{equation}\label{finaleqofstatepsi}
\sqrt{1-c^2z^8}\Big{(}-\psi''_{R/L}(z)+\frac{m_5(m_5\mp 1)-c^2z^8(2m^2_5+7)+c^4z^{16}m_5(m_5\pm 1)}{z^2(1-c^2z^8)^2}\psi_{R/L}(z)\Big{)}=M_n^2 \psi_{R/L}(z)
\end{equation}
In \eqref{finaleqofstatepsi}, $m_5$ is a parameter on the AdS side of gauge/ gravity duality and is related to the baryon mass on the gauge side, so the normalizable solutions of the  equations above are dual to the states in the boundary theory.
In pure AdS space, the bulk mass is related to the canonical conformal dimension $\Delta_{can}$ of a boundary operator as,
\begin{equation}\label{bulkmasspure}
|m^{AdS}_5|=\Delta_{can}-2.
\end{equation}
Recall that $QCD$ is not a conformal field theory since it has a mass gap. So the gravity side should be modified somehow and then it is not pure AdS any more. If one modifies AdS, the canonical dimension $\Delta_{can}$ of an
operator  has an anomalous contribution   $\gamma$  implying an effective scaling dimension.
\begin{equation}\label{bulkmass}
|m_5|=\Delta_{can}+\gamma-2.
\end{equation}
The contribution of the anomaly is related to how one modifies the theory. For example in \cite{Braga:2011wa}  modification of the scale introduces the mass gap in the theory. Therefore the anomalous contribution  represents the energy scale in the theory and leads to the  mass spectra. So, the main task is to find the value of the bulk mass in \eqref{bulkmass}. In AdS/CFT dictionary, the bulk mass is related to the dimension, means the energy scale of
the boundary theory is holographically related to the localization in the z-coordinate, therefore we have z-dependent mass in the bulk. Let us focus on $m_5$.  One can  fit $m_5$ numerically as the equations \eqref{finaleqofstatepsi} have normalizable solutions.
By fixing $M$ as  proton mass, we should find suitable values for $c$ and $m_5$ which give us well defined answers. \\
Figures \ref{chiinitial} and \ref{chi_final}  show initial (n=1) and final (for two excited states as n=2,3)  chiral components of the wave function respectively. To solve the equations \eqref{finaleqofstatepsi} numerically, we fix  proton mass $M$ as eigenvalue of equation, therefore $m_5$ and $c$ are found as  $c=0.0120  \, \rm GeV^4$ and $m_5=0.081 \, \rm GeV$. Interestingly, the value of parameter $c$ on the AdS side is very close to the phenomenological GC value of QCD as found $G_2 = 0.010 \pm 0.0023 \, \rm GeV^4$ in the reference \cite{ol070}. Another consequence of the presence of $c$ is that the anomaly $\gamma$ in \eqref{bulkmass} affects the bulk mass  intensely.
\begin{figure}[h!]
\begin{center}$
\begin{array}{cccc}
\includegraphics[width=7cm,height=5cm,clip]{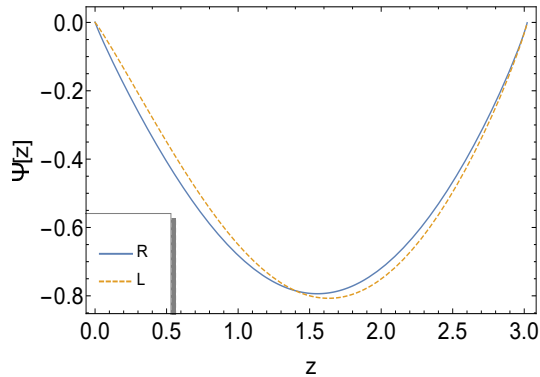}
\end{array}$
\end{center}
\caption{Left-handed (dashed) and right-handed (solid) sectors of wave function from \eqref{finaleqofstatepsi} for the initial state (target proton), by considering $c=0.0120 \, \rm GeV^4$ and $m_5=0.081 \,\rm GeV$ . }\label{chiinitial}
\end{figure}
\begin{figure}[h!]
\begin{minipage}[c]{1\textwidth}
\tiny{(a)}\includegraphics[width=7cm,height=5cm,clip]{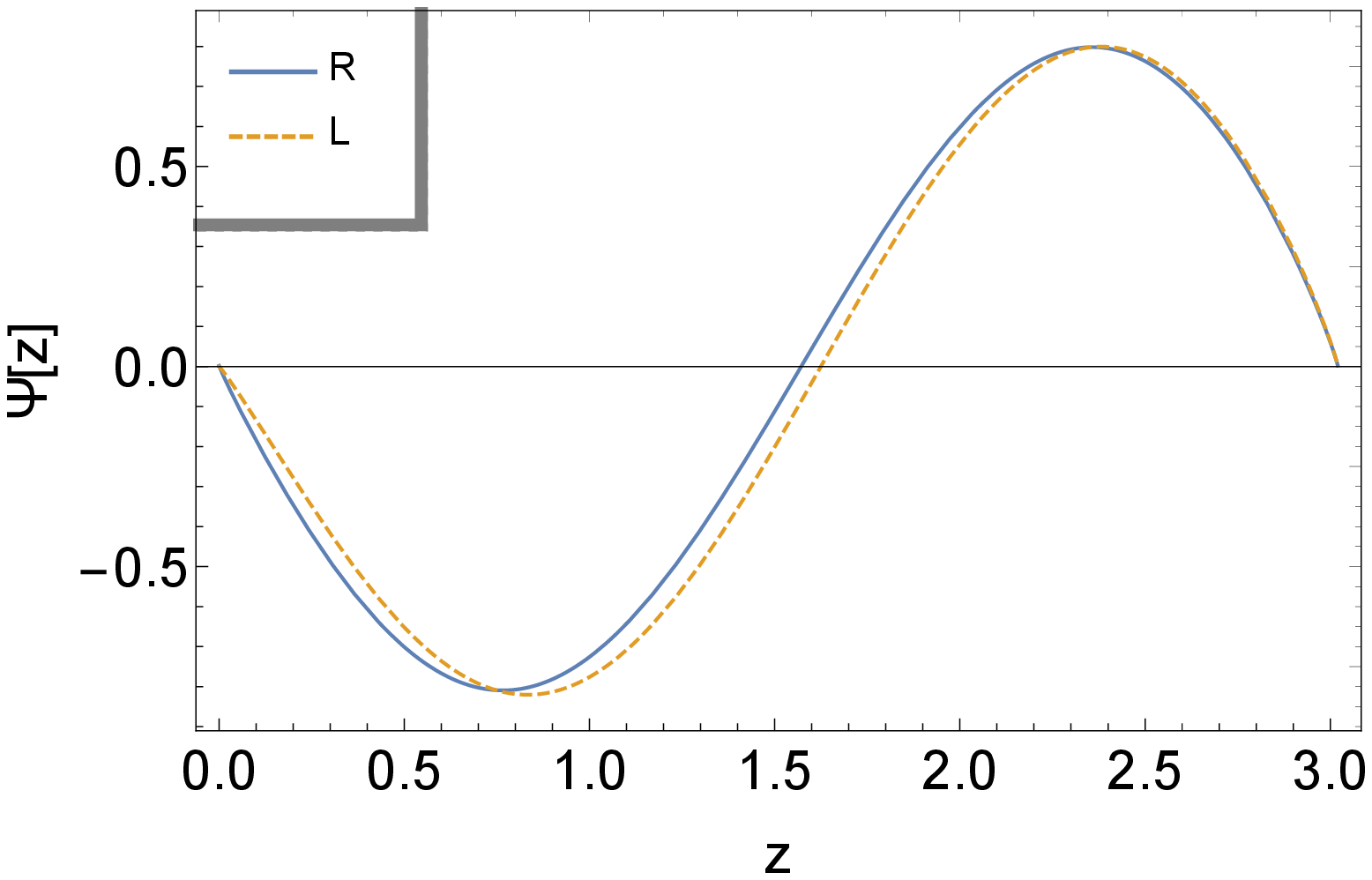}
\hspace{1cm}
\tiny{(b)}\includegraphics[width=7cm,height=5cm,clip]{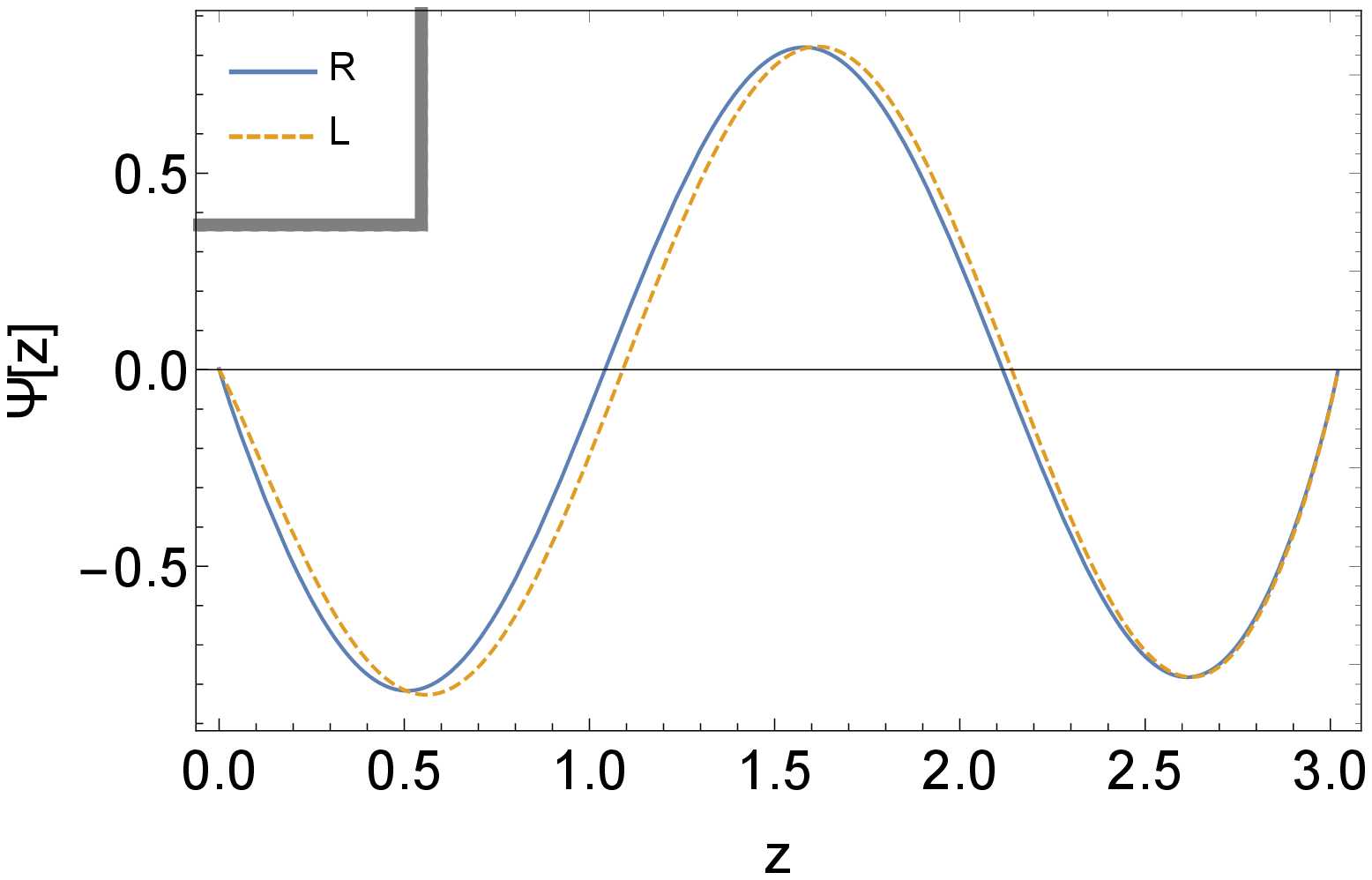}
\end{minipage}
\caption{Left-handed (dashed) and right-handed (solid) sectors of wave function from \eqref{finaleqofstatepsi} for the final state a)n=2 and b)n=3, by considering $c=0.0120  \,\rm GeV^4$ and $m_5=0.081\,\rm GeV$.}
\label{chi_final}
\end{figure}
After finding both left-handed and right-handed modes from \eqref{finaleqofstatepsi} we have,
\begin{equation}\label{initial spinor state}
\Psi_i=   \frac{e^{-\frac{2(1-c^2z^8)^{\frac{1}{4}}}{z}}}{(1-c^2z^8)^{\frac{1}{8}}}  e^{iP.y}[(\frac{1+\gamma_5}{2})\psi^i_L+(\frac{1-\gamma_5}{2})\psi^i_R] u_{s_i}(p),
\end{equation}
as the initial wave function for the target proton and,
\begin{equation}\label{final spinor state}
\Psi_X=   \frac{e^{-\frac{2(1-c^2z^8)^{\frac{1}{4}}}{z}}}{(1-c^2z^8)^{\frac{1}{8}}} e^{iP_X.y}[(\frac{1+\gamma_5}{2})\psi^X_L+(\frac{1-\gamma_5}{2})\psi^X_R] u_{s_X}(p),
\end{equation}
as the final wave function for the hadronic state. These will be used later.
\section{Modified geometry and the action of deep inelastic scattering}\label{se:sint}
According to  \eqref{eq:action and scattering} and \eqref{eq:action} we find the interaction action with the electromagnetic field and the baryonic states obtained from \eqref{secondeqofmofinal} and  \eqref{initial spinor state} -\eqref{final spinor state} respectively. The interaction action \eqref{eq:action} is written as,
\begin{eqnarray}\label{DISintaction}
S_{int}&=&g_{V}\int dz d^4y e^{-\varphi}\sqrt{-g} \phi^{\mu} \bar{\Psi}_{X} e^{\alpha}_{\mu} \gamma_{\alpha}\Psi_{i}\nonumber\\
&=&g_{V}\int dz d^4y \frac{1}{z}(1-cz^4)^{1+\sqrt{\frac{3}{2}}}(1+cz^4)^{1-\sqrt{\frac{3}{2}}}\phi^{\mu} \bar{\Psi}_{X} \frac{1}{z}(1-c^2z^8)^{\frac{1}{4}}\delta_{\mu}^{\alpha}\gamma_{\alpha}\Psi_{i}\nonumber\\
&=&g_{V}\int dz d^4y \frac{1}{z^2} (1-cz^4)^{\frac{5}{4}+\sqrt{\frac{3}{2}}} (1+cz^4)^{\frac{5}{4}-\sqrt{\frac{3}{2}}} \phi^{\mu} \bar{\Psi}_{X}\gamma_{\mu}\Psi_{i},
\end{eqnarray}
and from \eqref{final spinor state} one  writes,
\begin{equation}\label{final spinor statebar}
\bar{\Psi}_X=  \frac{e^{-\frac{2(1-c^2z^8)^{\frac{1}{4}}}{z}}}{(1-c^2z^8)^{\frac{1}{8}}} e^{-iP_X.y}\quad \bar{u}_{s_X}(p)[(\frac{1+\gamma_5}{2})\psi^X_L+(\frac{1-\gamma_5}{2})\psi^X_R].
\end{equation}
Therefore \eqref{DISintaction} is given by,
\begin{eqnarray}\label{DISintactionfinal}
S_{int}&=&\frac{g_{_{V}}}{2}\int dz d^4y e^{-i(P_{_X}-P-q).y} \eta^{\mu} \phi_1\frac{1}{z^2} (1-cz^4)^{\frac{5}{4}+\sqrt{\frac{3}{2}}} (1+cz^4)^{\frac{5}{4}-\sqrt{\frac{3}{2}}}\nonumber\\
 \frac{e^{-\frac{4(1-c^2z^8)^{\frac{1}{4}}}{z}}}{(1-c^2z^8)^{\frac{1}{4}}}&[&\bar{u}_{s_X}(\hat{P}_L\psi^{X}_{L}+\hat{P}_R\psi^{X}_{R})\gamma_{\mu}(\hat{P}_L\psi^{i}_{L}+\hat{P}_R\psi^{i}_{R})u_{s_i}]\nonumber\\
&=&\frac{g_{_{V}}}{2} (2\pi)^4\delta^4 (P_{_X}-P-q)  \eta^{\mu} \int dz \frac{1}{z^2} (1-cz^4)^{\frac{5}{4}+\sqrt{\frac{3}{2}}} (1+cz^4)^{\frac{5}{4}-\sqrt{\frac{3}{2}}}\nonumber\\
 \frac{e^{-\frac{4(1-c^2z^8)^{\frac{1}{4}}}{z}}}{(1-c^2z^8)^{\frac{1}{4}}}\phi_1 &[&\bar{u}_{s_X}\gamma_{\mu}\hat{P}_R u_{s_i}\psi^{X}_{L}\psi^{i}_{L}+\bar{u}_{s_X}\gamma_{\mu}\hat{P}_L u_{s_i}\psi^{X}_{R}\psi^{i}_{R}].\nonumber\\
\end{eqnarray}
By defining the following integral,
\begin{equation}\label{B term}
\mathcal{B}_{R,L}=\int dz  \frac{e^{-\frac{4(1-c^2z^8)^{\frac{1}{4}}}{z}}}{(1-c^2z^8)^{\frac{1}{4}}}\frac{1}{z^2} (1-cz^4)^{\frac{5}{4}+\sqrt{\frac{3}{2}}} (1+cz^4)^{\frac{5}{4}-\sqrt{\frac{3}{2}}} \phi_1\psi^{X}_{R,L}\psi^{i}_{R,L},
\end{equation}
\eqref{DISintactionfinal}  is written as,
\begin{equation}\label{BDISintactionfinal}
S_{int}=\frac{g_{_{V}}}{2}(2\pi)^4\delta^4 (P_{_X}-P-q)\eta^{\mu}[\bar{u}_{s_X}\gamma_{\mu}\hat{P}_R u_{s_i}\mathcal{B}_{L}+\bar{u}_{s_X}\gamma_{\mu}\hat{P}_L u_{s_i}\mathcal{B}_{R}],
\end{equation}
and \eqref{eq:action and scattering} is written as,
\begin{eqnarray}\label{final action and scattering}
\eta_{\mu}<P_X \vert J_{\mu}(q)\vert P_i>=\frac{g_{_{eff}}}{2}\delta^4 (P_{_X}-P-q)\eta_{\mu}[\bar{u}_{s_X}\gamma_{\mu}\hat{P}_R u_{s_i}\mathcal{B}_{L}+\bar{u}_{s_X}\gamma_{\mu}\hat{P}_L u_{s_i}\mathcal{B}_{R}]\nonumber\\
\eta_{\nu}<P_i \vert J_{\mu}(q)\vert P_X>=\frac{g_{_{eff}}}{2}\delta^4 (P_{_X}-P-q)\eta_{\nu}[\bar{u}_{s_i}\gamma_{\mu}\hat{P}_R u_{s_X}\mathcal{B}_{L}+\bar{u}_{s_i}\gamma_{\mu}\hat{P}_L u_{s_X}\mathcal{B}_{R}],
\end{eqnarray}
where $g^2_{_{eff}}=\mathcal{K}^2_{eff} g^2_{_{V}}(2\pi)^8$, and $g^2_{_{V}}=\frac{1}{137}$. $\mathcal{K}^2_{eff}$ should be fitted numerically as  shown in table 1.\\
With \eqref{hadronic tensor } we get,
\begin{eqnarray}\label{amp}
\eta_ \mu \eta_ \nu W^{\mu \nu}&=& \frac{\eta_{\mu \nu}}{4} \sum_{M_x^2}\sum_{s_i, s_X} \frac{g^2_{\rm eff}}{4}\,\delta(M^2_X-(P+q)^2) \left[\bar{u}_{s_X}\,\gamma^\mu\,\hat{P}_R\,u_{s_i}\, \bar{u}_{s_i}\,\gamma^ \nu\,\hat{P}_R\,u_{s_X}\,{\cal B}_L^2\right.\nonumber\\ &+&\left. \bar{u}_{s_X}\,\gamma^ \mu\,\hat{P}_R\,u_{s_i}\, \bar{u}_{s_i}\,\gamma^ \nu\,\hat{P}_L\,u_{s_X}\, {\cal B}_L\,{\cal B}_R\, + \,\bar{u}_{s_X}\,\gamma^ \mu\,\hat{P}_L\,u_{s_i}\,\bar{u}_{s_i}\,\gamma^\nu\,\hat{P}_R\,u_{s_X}\,{\cal B}_R\,{\cal B}_L \right. \nonumber\\ &+& \left. \bar{u}_{s_X}\,\gamma^ \mu\,\hat{P}_L\,u_{s_i}\,\bar{u}_{s_i}\,\gamma^\nu\,\hat{P}_L\,u_{s_X}\, {\cal B}_R^2\right]
\end{eqnarray}
Since our calculation is spin independent, we write,
\begin{equation}\label{spinless}
\sum_s  (u_s)_ \alpha (p) \, (\bar u_s)_ \beta (p) = (\gamma^ \mu p_ \mu + M)_{\alpha \beta},
\end{equation}
with the summation over the initial and final spin states, and
then applying trace engineering we arrive at,
\begin{multline}\label{NEWAMP}
\eta_ \mu \eta_ \nu W^{\mu \nu}=\frac{g^2_\text{eff}}{4}\,\sum_{M_X^2}\,\delta(M_X^2-(P+q)^2)\,\left\{(\mathcal{B}_L^2+\mathcal{B}_R^2)\left[(P\cdot \eta)^2-\frac{1}{2}\eta\cdot\eta(P^2+P\cdot q)\right]\right. \\
 +\mathcal{B}_L\,\mathcal{B}_R\,M_X^2\,M_0^2\,\eta\cdot\eta\biggr{\rbrace},
\end{multline}
where $\slashed p = \gamma^ \mu p_ \mu$,  $\{\gamma_5, \gamma_ \mu\} = 0$, and $P_{R/L} \gamma^ \mu = \gamma^ \mu P_{L/R}$.
 Summing over the outgoing states $P_X$ and carrying it on  to the continuum limit we have the invariant mass delta function
which is related to the functional form of the mass spectrum of the produced particles with the excitation number $n$ \cite{Polchinski:2002jw},
\begin{equation}
 \delta(M_{X}^2-(P+q)^2)\propto \left(\frac{\partial\,M^2_n}{\partial\,n}\right)^{-1}.
\end{equation}
We consider the lowest state produced at the collision, since the spectrum is linear with $n$  , and the delta will account for $1/ M_X^2$ \cite{Polchinski:2002jw, BallonBayona:2007qr}. With the  transversal polarization ($\eta\cdot q=0$),
the hadronic tensor \eqref{hadronic tensor } is obtained as,
\begin{equation}\label{finalW}
\eta_{\mu}\eta_{\nu} W^{\mu\nu}=\eta^{2} F_1(q^2,x)+\frac{2x}{q^2} (\eta.P)^2 F_2(q^2,x),
\end{equation}
where $F_1$ and $F_2$ are,
\begin{equation}\label{F1}
F_1(q^2,x)=\frac{g^2_{_{eff}}}{4}\Big{[}M_0M_X\mathcal{B}_{L}\mathcal{B}_{R}+(\mathcal{B}^2_{L}+\mathcal{B}^2_{R})(\frac{q^2}{4x}+\frac{M_0^2}{2})\Big{]}\frac{1}{M^2_X},
\end{equation}
and
\begin{equation}\label{F2}
F_2(q^2,x)=\frac{g^2_{_{eff}}}{8}\frac{q^2}{x}(\mathcal{B}^2_{L}+\mathcal{B}^2_{R})\frac{1}{M^2_X},
\end{equation}
respectively. Also $M_0$ is the mass of the initial
hadron and $M_X$ is the mass of the  final hadron as,
\begin{equation}
M_X=\sqrt{M_0^2+q^2(\frac{1-x}{x})}.
\end{equation}
\subsection*{Numerical strategy}\label{Numerical strategy}
As we mentioned in section \ref{se:Baryonic state's computation}, in the equation of states the eigenvalue of the ground state should  be close to the square of the  proton mass. So, we consider the ranges $0<m_5<1 \,\rm GeV$  and $0.001 <c <1\,\rm GeV^4$, while the  eigenvalue of the ground state equation, is in the range from $0.876\,\rm GeV$ to  $1\,\rm GeV$ ($M_{proton}=0.938 \, \rm GeV$). Accordingly we get a set of suitable values of the parameters $m_5$ and c. Their  approximate ranges are $0.001<m_5<0.2 \,\rm GeV$ and $0.006<c<0.02 \, \rm GeV^4$. In the next step we look for the appropriate values of $\mathcal{K}^2_{eff}$ as $m_5$ and $c$ satisfy their ranges and our theoretical calculations can be fitted with the experimental data for $F_2$. As it is shown in table 1, in accordance with the acceptable ranges for  $\mathcal{K}^2_{eff}$ , $m_5$ and $c$ we need to continue by  considering $0.01-0.04$ order  of $x$ and small $q^2$ which our model works well. In  sweep spectrum form, we determine a set of $m_5$ and $c$ and then for each set, we fit the experimental data for $F_2$ to get $ \mathcal{K}^2_{eff}$ by Least squares method. What needs to be mentioned here is that for $m_5$, the scanning step is 0.01, and for $c$, the scanning step is 0.001. The optimal  parameter values  with the smallest uncertainty  are $m_5=0.081 \,\rm GeV$, $c=0.0120 \, \rm GeV^4$, $\mathcal{K}^2_{eff}=37.3259 $. The uncertainty of $c$ comes from the size of scanning step. Using this   parameters set, we  will get the proton structure function $F_2$ as a function of $q^2$. \\

\begin{table*}[htb]
	\setcellgapes[t]{1.3mm}
	 \makegapedcells
	\setlength{\tabcolsep}{9mm}{
		\begin{tabular}{|c|c|c|c|}
		\cline{1-4}	
		$x$&$m_5/\rm GeV$&$c/\rm GeV^4$&$k^2_{eff}$\\ \cline{1-4}
			0.015&\multirow{3}{*}{0.081} &\multirow{3}{*}{0.012}&\multirow{3}{*}{37.3259}  \\ \cline{1-1}
			0.025& & &\\ \cline{1-1}
			0.04& & &\\ \cline{1-4}
		\end{tabular}}
	\caption{Adjustment of parameter $k^2_{eff}$  at different $x$ with $ c=0.0120 \,\rm GeV^4$.
 Remind that the value of parameter $c$ is demanded  by phenomenological value of proton mass. }
\end{table*}
\begin{figure}[h!]
\begin{minipage}[c]{1\textwidth}
\tiny{(a)}\includegraphics[width=4.5cm,height=5cm,clip]{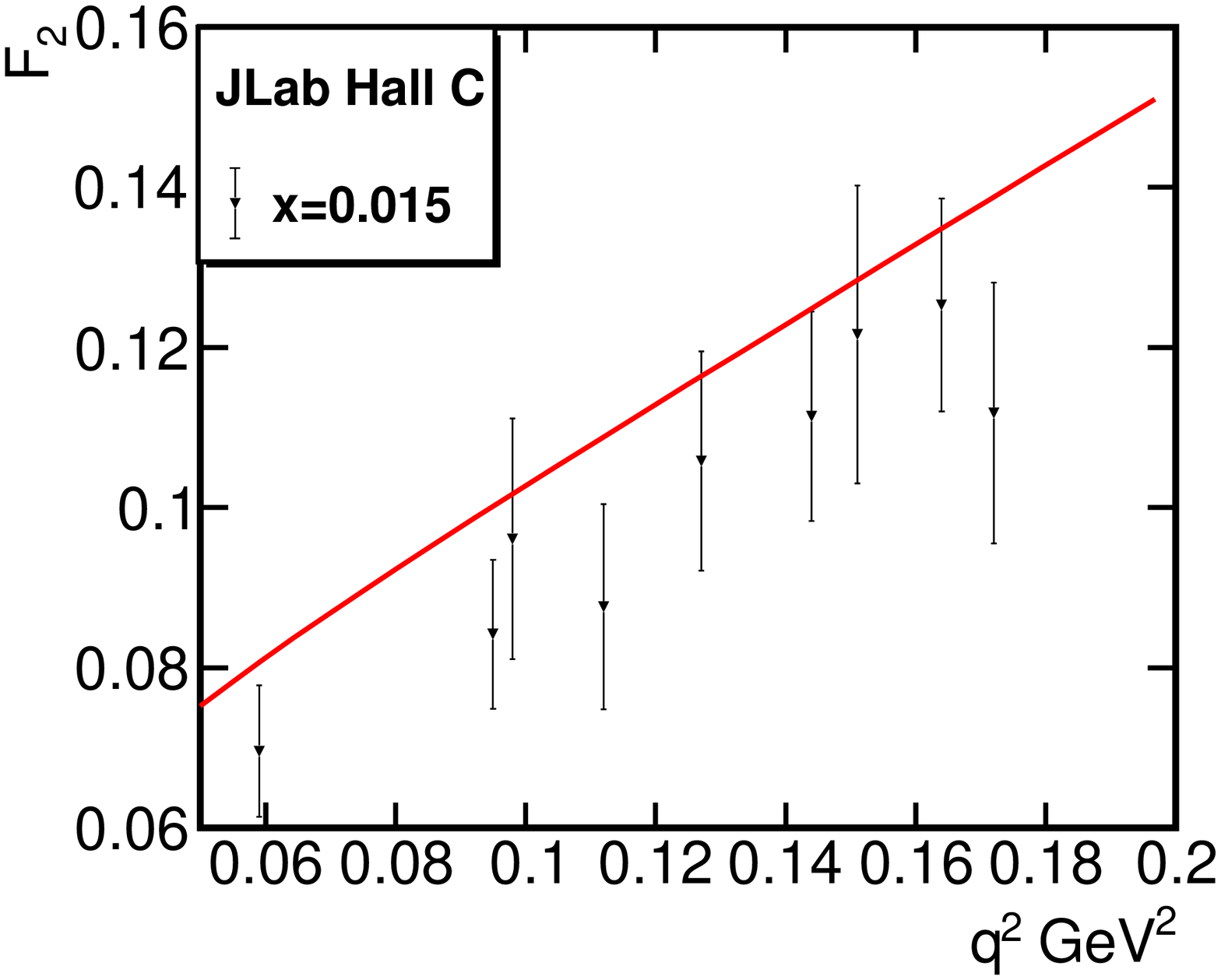}
\tiny{(b)}\includegraphics[width=4.5cm,height=5cm,clip]{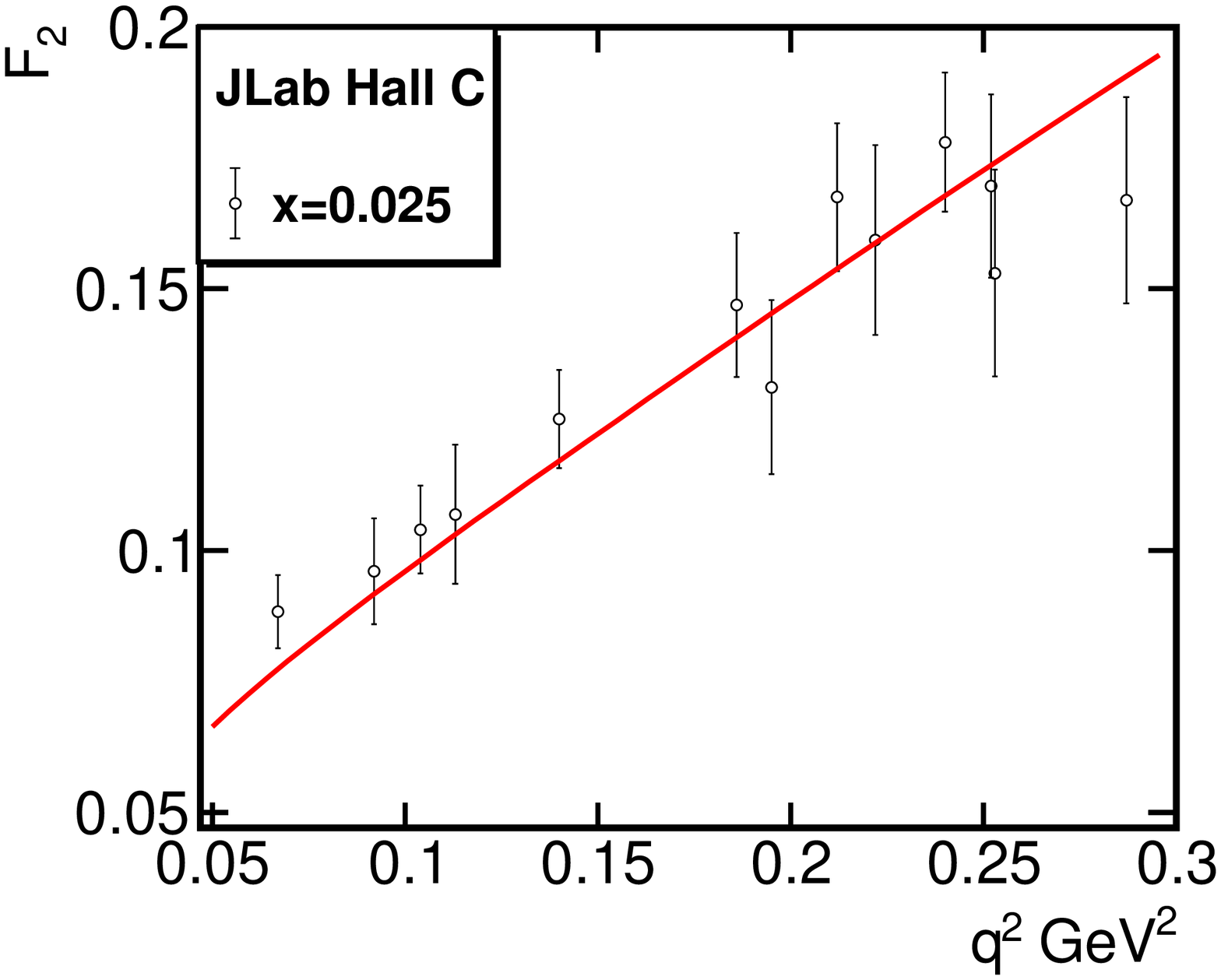}
\tiny{(c)}\includegraphics[width=4.5cm,height=5cm,clip]{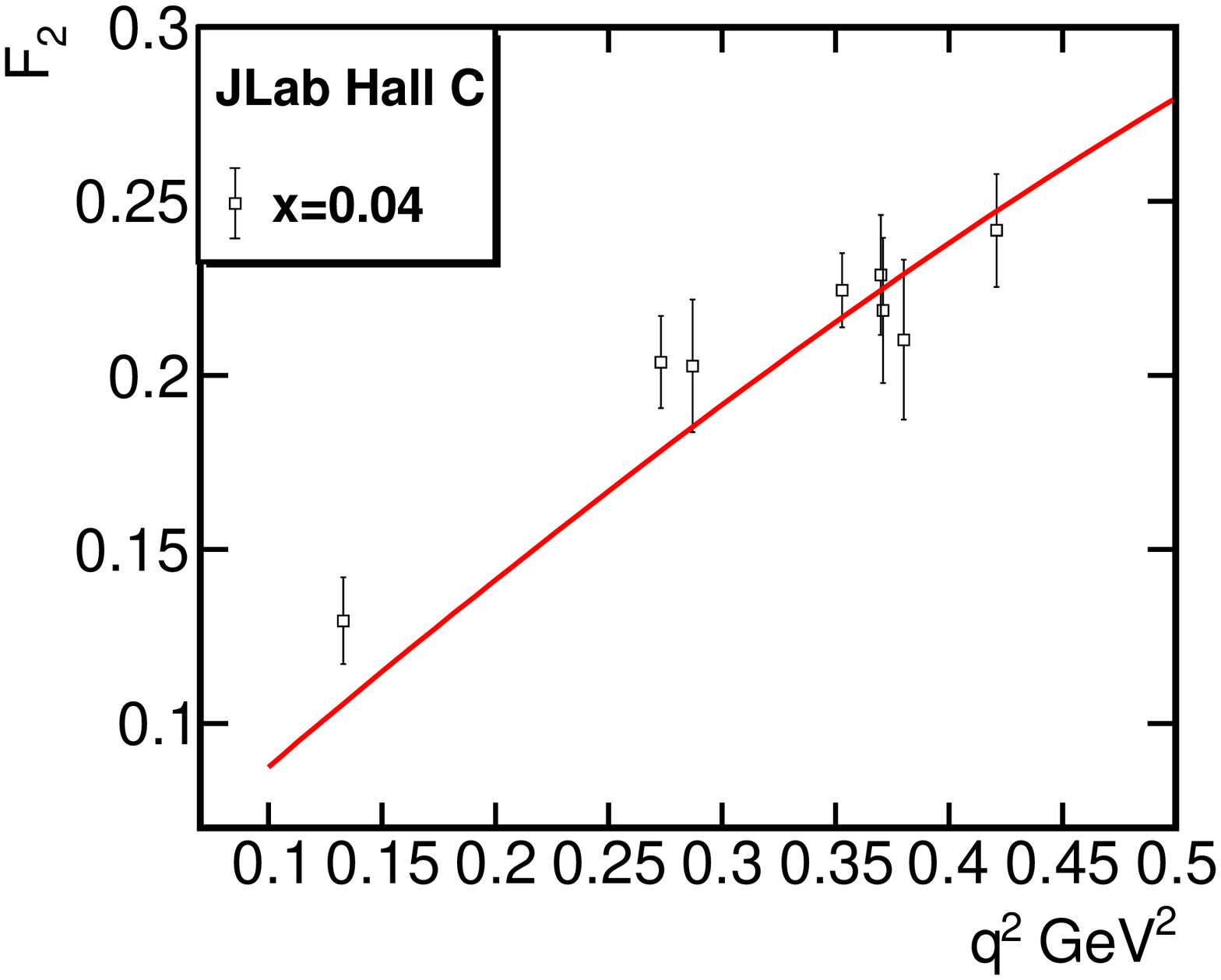}
\hspace{2.6cm}
\end{minipage}
\caption{a), b), c) Comparison between Jlab Hall C  data \cite{smallx } and our theoretical results. Dashed lines are theoretical results and square dots are experimental data.
}
\label{fig:Fsmallx}
\end{figure}
Figure \ref{fig:Fsmallx}   is a comparison between Jlab Hall C  data \cite{smallx } and our theoretical results. In plots a), b), c)  our results have  good agreements with experimental data at $x=0.015$, $x=0.025$ and $x=0.04$. 
Note that the electromagnetic wave function and the hadron function are independent of the Bjorken variable $x$. In our calculation this variable only appears in  the $F_2$ formula \eqref{F2} which introduces some uncertainty that is more visible in diagram (a).  In practice, the GC model is suitable for such  scattering for these values of $x$.  Applying the background \eqref{eq:metric} and \eqref{eq:dilaton} on the one hand and setting the proton as the DIS target on the other, results in  some specific small values of the $c$ parameter being acceptable, then our model  turned out to work better for the mentioned region of  $x$ values (which are not large),   than large values, as an alternative to \cite{coppro}  discussing large $x$ values  in the soft- wall model.

 There are differences between the applications of the SW model that was used in the same DIS in the \cite{coppro} and our current study. The modification parameter in \cite{coppro} appears in an exponential function having dimension mass squared $(GeV^2)$ and is associated with a QCD mass scale. The background metric is not coupled to a dilaton field.  Considering such  physics with the modified exponential function and due to the kinematical region in this reference their model is expected to produce better results for large $x$. With the goal of discussing other areas of $x$, we considered a different model. In our case the modification parameter introduced as $c$, has the dimension of $(GeV^4)$, also there is a coupled dilaton field in the action. The background structure and the proton target, strongly require the small value of the $c$ parameter and non-large $x$. 
\section{Conclusions}\label{sec:Conclusions}
In a holographic description of DIS we found effects of the parameter $c$ appearing  in the background metric representing gluon condensation  in the boundary theory. Since there is a proton target in the scattering, the mass of the proton and the value of $c$ parameter both play an important role in this study.
One of our main aim  was to determine the value of $c$ from experimental data.  First, we found the behaviour of the electromagnetic field with respect to $z$ for different values of $c$. We have shown that the $c$ parameter can increase  the magnitude of this field, especially for small values of $q^2$. Then, we solved  the  equation of baryonic wave function numerically and set the proton mass as the ground state eigenvalue to find the best  bulk mass, $c$ parameter and the $k^2_{eff}$ values .
 Hence, only small values of $c$ lead to a well-defined answer of the equation, or proton target requires a small value of $c$. 
It could be suggested that since the $c$ parameter  breaks the conformal symmetry, its value represents the confinement. 
So in our study case  confinement is not strong. \\
Based on the above results, we discussed the structure functions in the scattering versus Jlab Hall C data (with  order $0.01-0.04$ of $x$ and small $q^2$). Numerically,  we used an appropriate set of  bulk mass and $c$ in the form of a sweep spectrum which were already determined to fit experimental data and the structure function.  With this formalism at finite $x$ \cite{Polchinski:2002jw} our  results are useful for understanding QCD and proton structure.\\
\\
\textbf{Acknowledgement}\\
Authors would like to thank Danning Li, Zi-qiang Zhang and Wei kou for useful discussions. This work was supported by
Strategic Priority Research Program of Chinese Academy of Sciences ( XDB34030301).
ST is supported by the PIFI ( 2021PM0065).

\end{document}